\documentclass[preprint,12pt]{elsarticle}

\usepackage{amssymb}
\usepackage{slashed}
\usepackage{amsmath,amsfonts,mathrsfs}
\newcommand\g{\gamma}

\newcommand\e{\eta}

\newcommand\m{\mu}
\newcommand\n{\nu}

\renewcommand\r{\rho}

\renewcommand\j{\psi}

\renewcommand\d{\partial}

\newcommand{\no}{\nonumber}

\newcommand{\diracslash}[1]{#1\llap{/\kern2pt}}
\newcommand{\be}{\begin{equation}}
\newcommand{\ee}{\end{equation}}
\newcommand{\bea}{\begin{align}}
\newcommand{\eea}{\end{align}}
\newcommand{\ba}[1]{\begin{array}{#1}}
\newcommand{\ea}{\end{array}}

\begin{document}

\begin{frontmatter}



\title{A constrained theory of non-BCS type superconductivity in gapped Graphene}


\author{Vivek M. Vyas}
\author{Prasanta K. Panigrahi}
\address{Department of Physical Sciences, Indian Institute of Science Education \& Research (IISER) - Kolkata, Mohanpur, Nadia - 741252, India}

\begin{abstract}
We show that gapped Graphene, with a local constraint that current arising from the two valley fermions are exactly equal, shows a non-BCS type superconductivity. Unlike the conventional mechanisms, this superconductivity phenomenon does not require any pairing. We estimate the critical temperature for superconducting-to-normal transition via Berezinskii-Kosterlitz-Thouless mechanism, and find that it is proportional to the gap. 
\end{abstract}


\end{frontmatter}

% \linenumbers

%\section{}
%\label{}

\section{Introduction}

Graphene is an atom thick allotrope of Carbon. Owing to its hexagonal honeycomb lattice structure it has unusual electronic properties. As was first shown by Wallace, that the valence band and conduction band touch each other at six points in the momentum space, around which quasiparticle dispersion is linear \cite{wallace}. It was pointed out by Semenoff that, the low energy quasiparticle excitations around these points, out of which only two are inequivalent, actually satisfy the massless Dirac equation \cite{semenoff}. Although, Lorentz symmetry is not a true space time symmetry to start with, it emerges in the long wavelength limit because of honeycomb structure of the material, the velocity of light being replaced by Fermi velocity of the electrons. Experimental isolation of this material paved the way to simulate relativistic quantum physics and check validity of various fertile ideas \cite{novoselov}. The phenomena that make this material important include Klein paradox \cite{katsnelson},  room temperature quantum Hall effect \cite{qhe},  Andreev reflection \cite{beenakker} and universal conductance \cite{review2,rmp}. Fermion number fractionalization due to vortices \cite{jackiw1} and anyonic excitations \cite{jackiw2} have also been studied in this novel material.

Although, the experimental search for superconductivity in pure graphene is yet to be successful, possibility of superconductivity in graphene has attracted considerable attention from both theoretical and experimental sides \cite{super1,wilczek}. Experimental observation of proximity induced superconductivity in graphene, implies that graphene supports phase coherence \cite{indsup}. Variants of conventional pairing mechanisms are proposed for superconductivity in pure and doped graphene, and graphitic layers \cite{sonin,castroneto}. In Ref. \cite{baskaran}, the authors address the possibility of high $T_{c}$ superconductivity using resonating valence bond model for doped graphene. Doped graphite is known to be superconducting, and possible mechanisms giving rise to this effect have been proposed \cite{weller,super2}. In Ref. \cite{tanaka}, the authors propose a mechanism of superconductivity arising due the edge states in graphene. In Ref. \cite{shreecharan} authors address a possibility of non-pairing based superconductivity.

In the present paper, we show that Graphene with a finite non-zero band gap and a constraint that, the currents from both the valley fermions are always equal,  exhibits superconductivity. We find that Graphene with this constraint, possesses infinite DC conductivity, shows Meissner effect and flux quantisation. It is seen that, the Lagrange multiplier field introduced to implement the above constraint behaves like a Nambu-Goldstone mode of BCS theory, and plays the central role in realising superconductivity. However, unlike BCS theory, here Meissner effect and flux quantisation occur not due to Anderson-Higgs mechanism, but rather due to topological Chern-Simons coupling. We find that the full quantum theory has no propagating Dirac fermion, and only charge neutral fermion-hole bound pairs propagate. After a certain finite temperature, we observe that spontaneous proliferation of monopoles in Lagrange multiplier field takes place via Berezinskii-Kosterlitz-Thouless phase transition, which marks the superconductor-to-normal transition. In case of finite size Graphene with armchair egdes, we show existence of dissipationless chiral gapless edge modes, as a consequence of superconductivity in bulk, .     

The paper is organised as follows. In subsequent section, we discuss in detail both classical and quantum theory of Graphene with current constraint. A  study of electromagnetic response of this theory is done in section \ref{sec2}. Section \ref{sec3} deals with discussions about phase transition in this theory, followed by section \ref{sec4} where boundary theory is worked out. Section \ref{sec5} summarises the results and closes with a discussion about obtained results.

\section{\label{sec1}Current constrained model}

In tight binding approximation, the electron hopping Hamiltonian, defined on a hexagonal lattice with hopping energy $t$, reads:
\begin{equation}\label{h0}
H =-\sum_{\bf{r}} \sum_{\bf{i=1,2,3}} t \left (  a^{\dagger}({\bf{r}}) b({\bf{r}} + {\bf{s}_{i}}) + b^{\dagger}({\bf{r}}+{\bf{s}_{i}}) a({\bf{r}})  \right ), 
\end{equation}
where the fermion operators $a$ and $b$ act on the two interpenetrating sublattices A and B \footnote{Here vector {\textbf{r}} points to sublattice A, which is connected to B sublattice via vectors ${\bf{s}}_{1}=(1,\sqrt{3})\frac{l}{2}$, ${\bf{s}}_{2}=(1,-\sqrt{3})\frac{l}{2}$ and ${\bf{s}}_{3}=(-1,0)l$, with $l$ being the C-C bond length.}. As is well known, $H$ can be linearized around two Dirac points $K_{\pm}$ \footnote{In our convention ${\bf{K}}_{\pm} = ( \frac{2 \pi}{3 l}, \pm \frac{2 \pi}{3 \sqrt{3} l})$. }, to yield: 
\begin{eqnarray} \nonumber
\nonumber H_{D} = -i v_{F} \int d^{2}x \left\lbrace   \Psi^{\dagger}_{+}({\bf{r}}) {\bf{\sigma}} \cdot {\nabla} \Psi_{+}({\bf{r}}) \right. \\ \nonumber \left. + \Psi^{\dagger}_{-}({\bf{r}}) {\bf{\sigma^{\ast}}} \cdot {\nabla} \Psi_{-}({\bf{r}}) \right\rbrace,  
\end{eqnarray}
where $v_{F}= \frac{3 t l}{2}$ is the Fermi velocity (which is hence forth set to unity, along with $\hbar$), $\Psi^{\dagger}_{\pm}({\bf{r}}) = (a^{\dagger}_{\pm}({\bf{r}}),b^{\dagger}_{\pm}({\bf{r}}))$ and Pauli matrices are defined as ${\bf{\sigma}}=(\sigma_{x},\sigma_{y})$ \& ${\bf{\sigma^{\ast}}}=(\sigma_{x},-\sigma_{y})$ \cite{wallace,semenoff}. By selective doping of one sublattice or due to interaction with substrate, a local onsite potential can be generated which breaks sublattice symmetry, and leads to a mass gap in the electronic spectrum. This renders Graphene a semiconducting behaviour, and if mass gap is large enough then insulating nature follows.   
The Lagrangian describing these massive Dirac fermions can be written in a manifest Lorentz invariant form as:
\begin{equation} 
\mathscr{L}_{D} = \bar{\psi}_{+} ( i \gamma^{\mu}_{+}{\partial}_{\mu} - m  ) \psi_{+} + \bar{\psi}_{-} ( i \gamma^{\mu}_{-} {\partial}_{\mu} - m )\psi_{-}.
\end{equation}
Here, we have defined Gamma matrices for $\j_{+}$ field as $\g^{0}_{+} = \sigma_{3}, \g^{1}_{+} = i\sigma_{1} \text{  and  } \g^{2}_{+} = i\sigma_{2}$. Gamma matrices for $\j_{-}$ field are also same as $\j_{+}$ except for $\g^{2}$, which is defined as $\g^{2}_{+}=-\g^{2}_{-}$ (henceforth we shall use Feynman slash notation, whereby $\slashed{a}_{\pm}= \gamma^{\mu}_{\pm}a_{\mu}$). So we see that in Graphene, low energy electronic excitations are two species of Dirac fermions concentrated around two ($K_{\pm}$) valleys. We now consider a scenario where these quasiparticles are coupled to an interaction, which leads to preservation of valley symmetry locally. In what follows, we shall not be concerned about the detailed structure of this interaction, but make use of fact that, that currents generated in response to some external stimuli, from both valleys must exactly be the same locally $i.e.,$ $j^{\mu}_{+}(x,t)=j^{\mu}_{-}(x,t)$ ($j^{\mu}(x) = \bar{\psi}(x) \gamma^{\mu} \psi(x)$). Further, we shall also assume that, although the fermion fields are strictly interacting, the interaction allows above mentioned Dirac description of quasiparticles, and the net effect of interaction amounts only to the above local current constraint (LCC). 

The above condition endows the theory with a local gauge invariance. Indeed it is straightforward to check that, under following local gauge transformations:
\begin{align}\label{gtrans}
\psi_{+} & \rightarrow e^{-i \theta(x)} \psi_{+} \\
{\psi}_{-} & \rightarrow {\psi}_{-} e^{i \theta(x)}, 
\end{align}
the Lagrangian transforms as: 
\begin{equation}
\mathscr{L}_{D} \rightarrow \mathscr{L}_{D} + \left( j^{\mu}_{+} - j^{\mu}_{-} \right) \d_{\mu} \theta, 
\end{equation}
and since $j^{\mu}_{+} - j^{\mu}_{-} = 0$, the Lagrangian remains invariant under local gauge transformation. Hence, we have shown (at the classical level) that, the theory under consideration is an Abelian gauge theory, albeit without gauge field \cite{sri}. 

In functional integral formulation of quantum field theory, generating functional is an object of central importance, which for the above mentioned theory reads: 
\begin{align}  \label{z1}
& {Z} = N \int \mathscr{D}[\bar{\psi}_{(+,-)}, {\psi}_{(+,-)}]\, \delta \left( j^{\mu}_{+} - j^{\mu}_{-} \right) e^{i S_{D}}, \\
&\text{where}\,\,\, S_{D} = \int d^{3}x\,\mathscr{L}_{D}.
%j^{\mu} = \psi^{\dagger} \gamma^{\mu} \gamma^{3} \gamma^{5} \psi = 0.
\end{align}
Delta function in the above expression is introduced to implement LCC, and can be rewritten by introducing an additional Lagrange multiplier Bose field $a_{\mu}$, such that:
\begin{align}  \label{z2}
&Z[\bar{\psi}_{(+,-)},\psi_{(+,-)}] = N \int \mathscr{D} [\bar{\psi}_{(+,-)},{\psi}_{(+,-)},{a}_{\mu}] \, e^{i S_{D}[\bar{\j},\j,a_{\m}]}, 
\end{align}
where
\begin{align}
\no
S_{D}[\bar{\j},\j,a_{\m}] = \int d^{3}x \, & \left[ \bar{\psi}_{+}  ( i \slashed{\partial}_{+} - m + \slashed{a}_{+} ) \psi_{+} \right. 
\\ \label{lag} & \left. + \bar{\psi}_{-} ( i \slashed{\partial}_{-} - m - \slashed{a}_{-} ) \psi_{-} \right].
\end{align}
Modulo an unimportant normalisation constant, (\ref{z1}) and (\ref{z2}) describe same physics, as can be checked by integrating out $a_{\mu}$ field to reproduce the delta function constraint. This shows that above action captures dynamics of quasiparticles in Graphene subject to LCC. It is evident from the structure of the above action, that it remains invariant under local gauge transformations:
\begin{align} 
\psi_{+} \rightarrow e^{-i \theta(x)} \psi_{+},\,\, {\psi}_{-} \rightarrow {\psi}_{-} e^{i \theta(x)}, 
\,\, a_{\mu} \rightarrow a_{\mu} + \d_{\mu} \theta(x). \label{gtrans2}
\end{align}
Hence, one can identify $a_{\mu}$ as an Abelian gauge field, which is minimally coupled to two fermions, which are oppositely charged. It is easy to see that generating functional defined above is also invariant under above transformations, and no non-trivial Jacobian appears since these transformations are assumed to be regular. As a consequence, we expect that various $n$-point functions in this theory would obey Ward-Takahashi identities. In our case, two-point function for $K_{+}$ valley fermions, is given by:
\begin{align} \no
i S_{F}(x-y) & = \langle T \left( \j_{+}(x) \bar{\j}_{+}(y)  \right) \rangle \\  = N \int & \mathscr{D}[\bar{\psi}_{+},{\psi}_{+},{a}_{\mu}] \,\, \j_{+}(x) \bar{\j}_{+}(y) \,e^{i S_{D}[\bar{\j},\j,a_{\m}]}.   
\end{align}
Under field redefinition (\ref{gtrans2}), we see that propagator satisfies a Ward-Takahashi identity,
\begin{equation}
S_{F}(x-y)= e^{-i \theta(x)} S_{F}(x-y) e^{i \theta(y)},  
\end{equation}
whose only solution is $S_{F}(x-y) \propto \delta(x-y)$. Above identity is very powerful, since it has allowed for an exact determination of propagator in this interacting theory. Exactly similar identity would also hold for propagator of $K_{-}$ fermions. It is worth mentioning, that this model is one of the rare cases where  full propagator of this theory is known without any approximation. Presence of a physically observable particle in a theory, manifests as poles of propagator in momentum space. In our case, as is clearly evident, the propagator is regular everywhere in momentum space, which implies that Dirac fermion in our theory is not a propagating mode. This is particularly surprising since we started with a free Dirac theory with a constraint condition on currents, and it appears that condition is severe enough to not allow free fermion propagation.

In the absence of Dirac fermions, it is a natural to inquire about quasiparticle excitations in this theory. Inorder to answer this question, it is instructive to study the four-point function in this theory, which is defined as:
%\begin{widetext}
\begin{align} \no
\langle T &\left( \j_{+}(x_{1}) \j_{+}(x_{2}) \bar{\j}_{+}(y_{1}) \bar{\j}_{+}(y_{2})  \right) \rangle = \\ & N \int \mathscr{D}[\bar{\psi}_{+},{\psi}_{+},{a}_{\mu}] \,\, \j_{+}(x_{1}) \j_{+}(x_{2}) \bar{\j}_{+}(y_{1}) \bar{\j}_{+}(y_{2}) \,e^{i S_{D}}.   
\end{align}
%\end{widetext}
Under local gauge transformation (\ref{gtrans2}), we obtain another Ward-Takahashi identity for four-point function:
%\begin{widetext}
\begin{align} \no
\langle T & \left( \j_{+}(x_{1}) \j_{+}(x_{2}) \bar{\j}_{+}(y_{1}) \bar{\j}_{+}(y_{2})  \right) \rangle = \\
& e^{-i \theta(x_{1})} \, e^{-i \theta(x_{2})} \langle T \left( \j_{+}(x_{1}) \j_{+}(x_{2}) \bar{\j}_{+}(y_{1}) \bar{\j}_{+}(y_{2})  \right) \rangle \,\,e^{i \theta(y_{1})} \, e^{i \theta(y_{2})}. 
\end{align}
%\end{widetext}
   
Apart from a trivial non-propagating solution discussed above, assuming validity of translational invariance,  above equality admits a solution of the kind:
$\langle T \left( \j_{+}(x_{1}) \j_{+}(x_{2}) \bar{\j}_{+}(y_{1}) \bar{\j}_{+}(y_{2})  \right) \rangle \propto \delta{(x_{1}-y_{1})}\,\delta{(x_{2}-y_{2})}\,f(x_{1}-x_{2})$, where $f$ is some function of $(x_{1}-x_{2})$. This means that the above identity allows for propagation of composite operator $\j(x) \bar{\j}(y) |_{x=y}$, which describes charge neutral excitations consisting of fermion-antifermion bound states. In this case, these can be conveniently identified with exciton excitations of Graphene. Hence, we see that instead of Dirac fermion, Graphene with above constraint, admits excitons as its quasiparticle excitation. It is worth mentioning, that absence of fermions as elementary excitations and occurence of bound states in a constrainted theory like above, also appeared in a model of color confinement proposed by Rajasekaran and Srinivasan, and in related works \cite{sri,eguchi,kikkawa}. Interestingly, they showed that quarks and gluons (which appeared as bound states) did not propagate and were confined, whereas mesons (color neutral bound states of quarks) were propagating excitation in their model.

\section{\label{sec2}Electromagnetic response}

The external photon field, living in $3+1$ D space time, interacting with fermions confined on a plane can not be correctly described by $-\frac{1}{4}F_{\mu \nu} F^{\mu \nu}$ ($\mu$, $\nu=0,1,2$), since it has only one degree of freedom,  and not all components of physical photon field couple to matter. As shown in Ref. \cite{kovner}, one begins with action for electromagnetic field  $\int d^{4}x \frac{-1}{4}F_{\mu \nu} F^{\mu \nu} + j_{\mu} A^{\mu}$, where $j^{\mu}$ describes the matter current confined to XY-plane. Integrating out z-coordinate in the above action, and using Greens function identity, we get the three dimensional action as : $\int d^{3}x  \frac{-1}{4}F_{\mu \nu} \frac{1}{\sqrt{\partial^{2}}}F^{\mu \nu} + j_{\mu} A^{\mu} $, where $F_{\mu \nu}$ ($\mu$, $\nu=0,1,2$) describes 3D physical magnetic field $B_{z}$, in-plane components of electric field $E_{x}$ and $E_{y}$. So, Lagrangian (\ref{lag}) with suitable coupling to photon field reads:
\begin{align} \no \label{LagEM}
\mathscr{L} & = - \frac{1}{4 g^{2}} F_{\mu \nu} \frac{1}{\sqrt{\partial^{2}}} F^{\mu \nu} + \left[ \bar{\psi}_{+} ( i \slashed{\partial}_{+} - m + \slashed{a} + \slashed{A}  ) \psi_{+} \right. \\ & \left. + \bar{\psi}_{-} ( i \slashed{\partial}_{-} - m - \slashed{a} + \slashed{A} )\psi_{-} \right].
\end{align}
It needs to be pointed out here that, we have assumed velocity of light $c$ equal to the Fermi velocity $v_{F}$ which is set to unity. This is actually not the case, but as shown in Ref. \cite{dorey,kovner}, this does not change the qualitative behaviour of the theory. Also, notice that the electromagnetic coupling constant $g$ is not dimensionless, as in $QED_{4}$, but rather has dimension of square root of mass. Above Lagrangian describes our theory at tree (classical) level. Inorder to take into account effects due to quantum corrections, which arise from virtual fermion loop excitation, one needs to find out the effective action by integrating out fermion field. Fermion spectrum is gapped in our theory, and at low energies fermions are only excited virtually, hence integration of fermion fields is physically meaningful. Effective action upto quadratic terms in fields, obtained using derivative expansion of fermion determinant \cite{pkp1,das} reads:
\begin{equation}
{\mathscr{L}}_{eff} = - \frac{1}{6 {\pi} |m|} {f_{\mu \nu}} {f^{\mu \nu}} - \frac{ m} { \pi |m|} \epsilon^{\mu \nu \rho} A_{\mu} {\partial}_{\nu} {a}_{\rho},
\end{equation}
where an additional factor of $2$ has been multiplied to account for spin degeneracy of fermions. It can be shown that, in the limit of large $m$ this approximation is valid and higher order terms can be neglected.
As is evident, $a_{\m}$ did not have a kinetic term to start with, but fermion loops have made it dynamical, and indeed it can be identified as a genuine Abelian gauge field. Further, these two gauge fields are coupled by a mixed Chern-Simons term, which has a topological nature \cite{deser1,deser2,cs1,cs2,cs3}. In other words, this implies that $a_{\m}$ field has now become electromagnetically charged due to presence of virtual fermion cloud around it, with current being given by $J^{\mu}=\epsilon^{\mu \nu \rho} {\partial}_{\nu} {a}_{\rho}$. 
It is interesting to note that, by defining $A_{\pm} = A \pm a$ fields, the Chern-Simons term in above Lagrangian can be rewritten, modulo a constant, as : $ \epsilon_{\mu \nu \rho} A^{\mu}_{+} {\partial}^{\nu} {A}^{\rho}_{+} - \epsilon_{\mu \nu \rho} A^{\mu}_{-} {\partial}^{\nu} {A}^{\rho}_{-}$, which has two topological terms with opposite signs; as a result it does not violate parity. Hence, the net action describing interaction of Graphene with electromagnetic field is given by:  
\begin{align} \nonumber
S = \int d^{3}x &\,\mathscr{L}, \,\, \text{where}\\
\mathscr{L} = - \frac{1}{4} F_{\mu \nu} \frac{1}{\sqrt{\partial^{2}}} F^{\mu \nu} & - \frac{1}{6 {\pi} |m|} {f_{\mu \nu}} {f^{\mu \nu}} - \frac{ gm} { \pi |m|} \epsilon^{\mu \nu \rho} A_{\mu} {\partial}_{\nu} {a}_{\rho} .
\end{align}
Note that, above action is invariant under two types of local gauge transformation: $a_{\mu} \rightarrow a_{\mu} + \d_{\mu} \lambda$ and $A_{\mu} \rightarrow A_{\mu} + \d_{\mu} \chi$, where $\lambda$ and $\chi$ are some regular functions of $x$, and we have assumed that fields under discussion decay sufficiently quickly as one approaches Graphene boundary. In general, it is easy to show, that these gauge invariance would imply presence of a massless mode living on Graphene boundary. However, in what follows, we shall assume that Graphene sheet is practically infinite, and discussion about bulk-boundary interaction will be pursued later. Further, first two terms in above Lagrangian are invariant under these transformations, whereas the third Chern-Simons term is not. Inorder to appreciate consequences arising due to this noninvariance, we follow Ref. \cite{dorey} and do a Wick rotation: $t \rightarrow - i \tau$ (where $\tau \in [0,\beta]$ is a compact variable), from Minkowskii space-time to Euclidean space-time. In this compact Euclidean space-time, bosonic (fermionic) fields obey periodic (anti-periodic) boundary conditions: 
\begin{align}
&\text{Bosonic:}\, B(\vec{x},0)=B(\vec{x},\beta), \\ & \text{Fermionic:}\, \j(\vec{x},0)=-\j(\vec{x},\beta).
\end{align}
These conditions alongwith requirement of single valuedness, impose a restriction on gauge function $\lambda(x)$: $\lambda(\beta) = \lambda(0) + {2 \pi n}$, ($n$ being integer, often called the winding number). Vacuum functional is defined as:
\begin{align} \label{zeuclid}
Z^{Euclid} = & N \int \mathscr{D} A_{\mu} \mathscr{D} a_{\mu} \, e^{- S_{CS} }, \,\, \text{where}\\
S_{CS} = & \int_{0}^{\beta} d \tau \int d^{2}x \,\, \frac{gm}{\pi |m|} \epsilon^{\mu \nu \rho} A_{\mu} {\partial}_{\nu} {a}_{\rho}.
\end{align}
Considering variation of Chern-Simons action under restricted gauge transformations where $\lambda(\tau)$ only depends on $\tau$, one finds:
\begin{align}
\delta S_{CS} &= \frac{2gm}{|m|}n \Phi, \,\, \text{where}\\
\Phi & = \int d^{2}x \, \epsilon_{ij} \d_{i}A_{j}, 
\end{align}
is magnetic flux. Invariance of vacuum functional (\ref{zeuclid}) under above transformation, demands:
$\delta S_{CS} = 2 \pi i N$, where $N$ is an integer. This clearly implies that $\Phi = \frac{N \pi |m|}{g m}$, which when written in SI units reads:
\begin{align}
\Phi = N \left(\frac{m}{|m|} \right) \frac{h c }{2 g}.
\end{align}
Hence, magnetic flux is quantised in this model with flux unit $\frac{h c }{2 g}$.

Inorder to understand response of the system due to influence of external electromagnetic field, we integrate out $a_{\mu}$ field from  Lagrangian (\ref{LagEM}), to arrive at an effective action for electromagnetic field:
\begin{align}
\mathscr{L}_{eff} = \frac{3 |m|}{4 \pi} \left( A_{\m} A^{\m}  - A_{\m} \frac{\d^{\m} \d^{\n}}{\d^{2}} A_{\n}  \right).
\end{align}
As is evident, interaction with $a_{\m}$ field has induced mass $M = \tfrac{3 |m|}{4 \pi}$ for the physical electromagnetic field. The photon field becoming massive implies Meissner effect, a hallmark of superconductivity, where the static magnetic field exponentially dies down with distance from the boundary, with characteristic length scale, called penetration depth, which in our case is $\lambda = \tfrac{4 \pi}{3 |m|}$. Further, current-current correlation function in this theory, can be found from above Lagrangian to be:
\begin{align} \nonumber
\langle j^{\m}(x) j^{\n}(y)\rangle &= \frac{\delta^{2} S_{eff}}{\delta A_{\m}(x) \delta A_{\n}(y)} \\
&= \frac{3 |m|}{2 \pi} \left( \eta^{\m \n} - \frac{\d^{\m} \d^{\n}}{\d^{2}} \right) \delta^{3}(x-y).
\end{align}
Notice, that above function has a pole at zero momentum, which implies infinite DC conductivity from Kubo formula. To make the comparison with usual BCS theory clear, we follow the technique given in Ref. \cite{banks}, and define an auxillary scalar field $\phi$, such that $\partial_{\mu} \phi = \epsilon_{\mu \nu \rho} \partial^{\nu} a^{\rho}$, so that the above Lagrangian after a Hubbard-Stratonovich transformation, reads:
\begin{equation}
{\mathscr{L}}_{eff} =  \frac{3 {\pi} |m|}{2} \left( {\partial}_{\mu} \phi + \frac{ m}{2 \pi |m|} A_{\mu}  \right)^{2}. 
\end{equation}
Inorder to make comparison with Landau-Ginzburg theory apparent, one can define $\chi = e^{i \phi}$, so that the above Lagrangian reads:
\begin{equation}
{\mathscr{L}}_{eff} =  \frac{3 {\pi} |m|}{2} \left| \left( {\partial}_{\mu} +  i \frac{ m}{2 \pi |m|} A_{\mu}  \right) \chi \right|^{2}. 
\end{equation}  

As is clear, the above Lagrangian is in manifest London form, and all the phenomenological properties of superconductivity would follow from here \cite{wein}. It was shown in generality by Weinberg \cite{wein}, that occurrence of an electromagnetically charged field $\xi(x)$, which transforms as $\xi(x) \rightarrow \xi(x) + \Lambda(x)$ under a gauge transformation, is sufficient for the theory to exhibit superconductivity. In case of theories which exhibit any kind of fermion pairing, phenomenon of spontaneous symmetry breaking takes place, whereby system realises a vacuum, which is not invariant under all symmetry group transformations that preserve the Hamiltonian. If the symmetry not respected by vacuum, is local gauge invariance of electrodynamics, then there inevitably appears a massless Nambu-Goldstone field, which transforms exactly like $\xi(x) \rightarrow \xi(x) + \Lambda(x)$ under a gauge transformation \cite{Nambu,Goldstone}. As argued by Weinberg, it is presence of this Goldstone mode that is ultimately responsible for superconductivity. One can see from above Lagrangian, that in our theory, $\phi$ field behaves exactly like the Nambu-Goldstone mode, and hence our theory exhibits genuine superconductivity. However, it is worth noting, that in our theory, the origin of $\phi$ field is due to a local constraint, and not due to breaking of any local symmetry by vacuum. In other words, our theory exhibits superconductivity due to presence of a local constraint, rather than that of fermion pairing. Also it is worth pointing out that, this non-BCS type superconductivity relies on presence of Chern-Simons term and hence is present only in the planar $2+1$ dimensional world.

\section{\label{sec3}Phase transition}

In above discussion, we have tactically assumed that $a_{\m}$ field is regular and single valued everywhere. 
In general, this may not be the case, which allows for occurrence of monopoles corresponding to $a_{\m}$ field, $i.e.,\, \epsilon_{\m \n \rho} \d^{\m} f^{\n \rho} \neq 0$. Equivalently this translates to presence of vortices in $\phi$ field, which means:
\begin{align}
\oint_{C} \vec{\nabla} \phi \cdot d\vec{s} = \pm 2 \pi n, \,\,\ (n \,\, \text{being an integer})
\end{align}
along any closed curve $C$ that enclosed the vortex. Interestingly, in absence of $A_{\m}$ field, above action exactly matches with that of continuum limit of 2D XY model. In case of XY model, it was shown by Berezinskii \cite{ber} and independently by Kosterlitz and Thouless \cite{kt} that, presence of vortex excitation is forbidden energetically at zero temperature, since it does not correspond to minimum of free energy. At zero temperature, vortices and antivortices are bound together due to mutual attractive potential, and hence do not move freely. However, at temperatures above a certain critical temperature, they showed that, presence of vortices is energetically favourable as it minimizes free energy. This means that at critical temperature, a transition from bound vortex state to free vortex state takes place, which is known as Berezinskii-Kosterlitz-Thouless (BKT) phase transition. This is an infinite order phase transition, wherein the ordered(low temperature) phase differs from disordered(high temperature) phase in terms of topology of field, rather than symmetry. It was shown that presence of these vortices, destroys the long range correlation present otherwise in the system. It is also known that the ordered phase in case of XY model, does not exhibit any Long Range Order, but shows what is called a Quasi Long Range Order, whereby the correlation function exhibits a power law decay asymptotically. So as in case of XY model, in this case also one expects a BKT phase transition to take place, after which free vortex excitations could be present. The critical temperature for this transition, can be read off from above action, which is given by 
\begin{equation}
T_{BKT} = \frac{3 \pi^{2} |m| }{2}.
\end{equation}
Note, that the critical temperature depends linearly on mass gap, and for gapless fermionic spectrum, the transition would occur at zero temperature.

The effect of vortices on electromagnetic response can be understood by writing an effective action for electromagnetic field after taking into consideration occurrence of vortices. Following Ref. \cite{Mackenzie1,Mackenzie2}, we take vortex excitation into account, in above action by writing $\phi \rightarrow \phi_{reg} + \phi_{vor}$, where $\phi_{reg(vor)}$ is regular(vortex) part of $\phi$, so that action reads:
\begin{equation}
{\mathscr{L}}_{eff} =  \frac{3 {\pi} |m|}{2} \left( {\partial}_{\mu} \phi_{reg} + {\partial}_{\mu} \phi_{vor} + \frac{ m}{2 \pi |m|} A_{\mu}  \right)^{2}. 
\end{equation}
Using an auxiliary field $\xi_{\m}$, above can be rewritten as:
\begin{equation}
{\mathscr{L}}_{eff} = - \frac{\xi^{2}}{3 \pi |m|} + \xi_{\m} \d^{\m} \phi_{reg} + \xi_{\m} \d^{\m} \phi_{vor} + \frac{\text{sgn(m)}}{2 \pi} \xi_{\m}A^{\m}.
\end{equation}
Integrating out regular part of $\phi$, implies a constraint: $\d^{\m}\xi_{\m} =0$, which has an obvious solution $\xi_{\m} = \epsilon_{\m \n \r} \d^{\n} B^{\r}$, where $B_{\m}$ is some gauge field. Above action written in terms of $B$ field is given by:
\begin{align} \no
{\mathscr{L}}_{eff} = & \frac{1}{3 \pi |m|} B_{\m}(\e^{\m \n} \d^{2} - \d^{\m} \d^{\n}) B_{\n} \\
& + B_{\m} \left( \epsilon^{\m \n \r} \d_{\n} \d_{\r} \phi_{vor} + \frac{sgn(m)}{2 \pi} \epsilon^{\m \n \r} \d_{\n} A_{\r} \right).
\end{align}
Integrating out auxiliary $B$ field, one gets an effective action describing interaction of vortex current $K^{\m} = \epsilon^{\m \n \r} \d_{\n} \d_{\r} \phi_{vor}$ with electromagnetic field $A_{\m}$:
%\begin{widetext}
\begin{align} \no
{\mathscr{L}}_{eff} = & - \frac{3 \pi |m|}{4} \left[ K^{\m}\frac{1}{\d^{2}}K_{\m} + \frac{sgn(m)}{2 \pi} \epsilon^{\m \n \r} \d_{\n} A_{\r} \frac{1}{\d^{2}} K_{\m} \right. \\ & \left. + \frac{sgn(m)}{2 \pi} \epsilon^{\m \n \r} K_{\m} \frac{1}{\d^{2}} \d_{\n} A_{\r} + \frac{1}{4 \pi^{2}} A_{\m} \left( \frac{\d^{\m} \d^{\n}}{\d^{2}} - \e^{\m \n} \right) A_{\n} \right].
\end{align}
%\end{widetext}
Notice, that the last term in above expression, is topologically trivial and is responsible for superconductivity. In absence of vortex current, it solely describes the response of this system. First term on the otherhand, describes interaction between vortices, and it can be easily shown that interaction potential between them has a logarithmic behaviour. The other two terms, in the effective action, describe interaction of vortices with electromagnetic field, which means that vortices in our model are charged. So, in order to find contribution of these charged vortices to electromagnetic response, one requires to integrate out vortex field, to yield a net effective action for electromagnetic field. Very interestingly, integration over vortex current contributes a term of the form $A_{\m} \left( \frac{\d^{\m} \d^{\n}}{\d^{2}} - \e^{\m \n} \right) A_{\n}$ which exactly cancels with the one already present in above action. This straightforwardly means that, in presence of vortices, superconductivity is destroyed \cite{Mackenzie2}. Hence, in this model, we see that occurrence of BKT phase transition marks a superconducting-to-normal phase transition, and therefore Graphene with local current constraint, at sufficiently low temperatures, exhibits superconductivity with strong type-II character.

\section{\label{sec4}Boundary theory}
  
As mentioned earlier, we have assumed that Graphene sheet is of infinite extent, and fields fall of sufficiently quickly, so that surface terms give negligible contribution. However, that is rarely the case in reality, where in general one encounters Graphene samples with boundary. Owing to its hexagonal tiling, Graphene can exhibit boundary of two kinds: Arm chair and Zig-zag. It is known that, the latter exhibits localised electronic egde states, whereas the former does not. Hence, in case of arm-chair egdes, fermions present in bulk can freely interact with the ones living on boundary and vice versa. In what follows, we shall assume that Graphene sheet has a well defined regular boundary of arm-chair kind.

As noted above, low energy effective action describing dynamics of low energy electronic excitation, subject to the local constraint, coupled to electromagnetic field is given by: 
\begin{align} \nonumber
\mathscr{L} = - \frac{1}{6 {\pi} |m|} {f_{\mu \nu}} {f^{\mu \nu}} - \frac{ m} { \pi |m|} \epsilon^{\mu \nu \rho} A_{\mu} {\partial}_{\nu} {a}_{\rho}.
\end{align}
As was observed earlier, the last term in above Lagrangian is not invariant under local gauge transformation: 
$a_{\mu} \rightarrow a_{\mu} + \d_{\mu} \Lambda$, where $\Lambda$ is some regular function of $x$. As a result, the change in action is given by:
\begin{equation}
\delta S_{CS} = \left( \frac{sgn(m)}{2 \pi} \right) \int d^{3}x \,\, \epsilon^{\m \n \r} \d_{\m}
 \left( \Lambda \, f_{\n \r} \right). 
\end{equation}
Above volume integral can be converted to a surface integral, defined on closed Graphene boundary, to give an action:
\begin{equation}
\delta S_{CS} = \left( \frac{sgn(m)}{2 \pi} \right) \int_{B} d^{2}x \,\, \epsilon^{\m \n } \Lambda \, f_{\m \n}.  
\end{equation}
This term, as it stands, is not gauge invariant, and is defined on Graphene boundary, which encloses the bulk. Gauge invariance of any given theory, is a statement that, the theory is constrained, and possesses redundant variables. We observe that, our theory to start with was gauge invariant at classical level. One loop corrections arising out of fermion loops, generate Chern-Simons term, which exhibits gauge noninvariance. Because, our theory to start with was gauge invariant, and hence constrained, consistency demands that quantum(corrected) theory should also respect the imposed constraints, and hence should be gauge invariant. The occurrence of above gauge noninvariance, simply implies that one is only looking at one particular sector of theory, and there exists other dynamical sector, whose dynamics is such that it compensates with the one above to render the total theory gauge invariance. Following Ref. \cite{wilczekbook}, we demand that there must exist a corresponding gauge theory living on the boundary, defined such that it contributes a gauge noninvariant term of exactly opposite character and hence cancels the one written above. The simplest term, living on boundary, that obeys above condition is:
\begin{equation}
S_{B} = \frac{-sgn(m)}{2 \pi} \int_{B} d^{2}x \, \theta \epsilon^{\m \n} f_{\m \n},
\end{equation}
where $\theta(x,t)$ is St\"uckelberg field, which transforms like $\theta \rightarrow \theta + \Lambda$.  
In general, this scalar field would be dynamical, and with a gauge invariant kinetic term, the boundary action reads:
\begin{equation}
S_{B} = \int_{B}d^{2}x \,\, \left[ c \,\left( \d_{\m} \theta - A_{\m} \right)^{2} - \frac{sgn(m)}{2 \pi} \theta \epsilon^{\m \n} f_{\m \n} \right].
\end{equation}
Note, that because of its peculiar transformation property, a quadratic mass term for $\theta$ is not gauge invariant. Hence, in a gauge theory framework like this, $\theta$ field remains massless. In deriving above action, we have only considered gauge invariance with respect to transformation in $a_{\m}$ field. However, analogously the same may be done for $A_{\m}$ field, so that net action, describing massless surface modes, coupled to both gauge fields is given by:
\begin{align}
S_{B} = \int_{B}d^{2}x \,\, & \left[ c \,\left( \d_{\m} \theta - a_{\m} - A_{\m} \right)^{2} \right. \\ &
\left. - \frac{sgn(m)}{2 \pi} \theta \epsilon^{\m \n}\left( f_{\m \n} + F_{\m \n} \right) \right].
\end{align}
There are several things to note here. Firstly, the action for $\theta$ is in manifest London form, and hence is indicative of non-dissipative transport on the boundary. Secondly, $\theta$ field is electromagnetically charged, and hence boundary supports dissipationless electric current, or in other words boundary is superconducting. Thirdly, the coupling of $\theta$ field, with that of $a$ and $A$ field is anomalous, as a result chiral current in this quantum theory is no longer conserved. This ultimately results in chirality of these surface modes. This can be explicitly seen by observing that, unlike bulk fermionic coupling, these surface modes couple to sum of two gauge fields $a+A$, and not to difference $a-A$. Hence, we have shown that, arm chair edged Graphene with LCC possesses dissipationless chiral gapless surface modes.  

\begin{center}  
\begin{figure}
\includegraphics[scale=0.5]{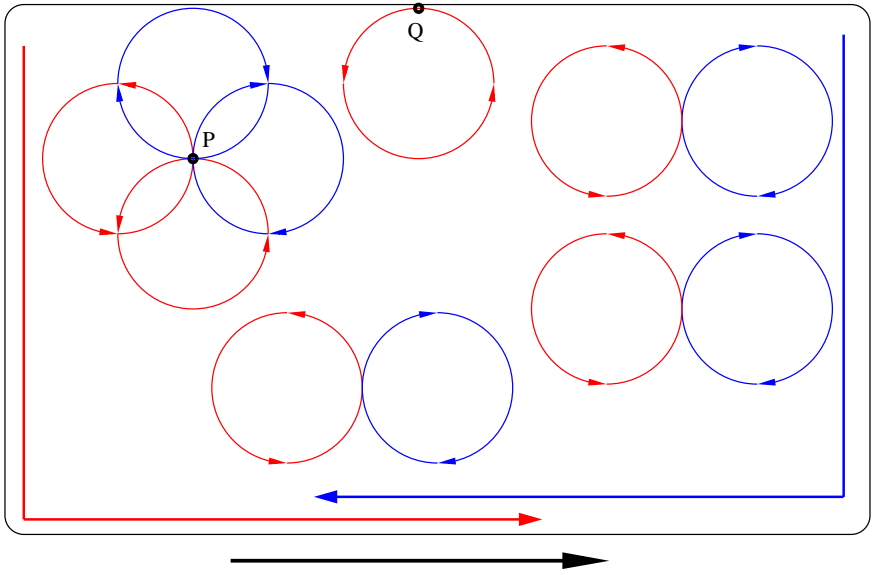}
\caption{\label{fig} Figure depicting motion of $K_{+(-)}$ valley fermions in red (blue), in Graphene under LCC. LCC is maintained at any point $P$ in  bulk, as it gets equal contribution from both the valleys in all directions. Points on the boundary,  however, like point $Q$, get contribution only from Graphene interior, which leads to existence of clockwise and counter clockwise edge currents (shown by red and blue arrows), one from each valley, such that the net edge current adds up to zero. But, Graphene edge being one dimensional, possesses chiral anomaly, which gives rise to a net non-zero chiral edge current (shown by black arrow).}
\end{figure}
\end{center}

\section{\label{sec5}Conclusion}
We have shown that, gapped Graphene with a local constraint $j^{\m}_{+}=j^{\m}_{-}$, exhibits superconductivity, albeit without any pairing. 
Apart from infinite DC conductivity, we show that this theory also realises Meissner effect, a hallmark of superconductivity. However it is worth noting that, unlike BCS theory, here Meissner effect does not occur due to Anderson-Higgs mechanism, whereby photon becomes massive by absorbing gapless Nambu-Goldstone mode. Rather, the origin of Meissner effect (or presence of photon mass) is due to presence of topological Chern-Simons term in the theory. As is known, Chern-Simons term is unique to two spatial dimensions, which means that this type of superconductivity only occurs in planar world and its higher dimensional extension may not be possible. It is seen that, the Lagrange multiplier field introduced to implement above local constraint, is found to behave analogous to Nambu-Goldstone mode of BCS theory. In pairing based theories, because of spontaneous breaking of symmetry there appears an excitation mode called the Amplitude mode \cite{umezawa}. In our theory, there is no fermion pairing, which implies absence of amplitude mode. This feature makes 
this superconductivity distinguishable from the other pairing based theories. At a critical nonzero temperature, like in 2D XY model, we see that spontaneous proliferation of monopoles in Lagrange multiplier field takes place, and marks occurrence of BKT phase transition. It is seen that the contribution  due to these monopole (singular) part of this field, exactly cancels the same coming from the regular part. So loss of superconductivity takes place via BKT mechanism, and superconductor-to-normal transition takes place at BKT critical temperature, which is proportional to mass gap, and hence can be controlled. On a finite graphene sheet with armchair egdes, we show that, this type of superconductivity naturally supports gapless chiral edge modes. Amusingly, we find that the full quantum theory does not have Dirac fermions as propagating mode, and charge neutral fermion-hole bound pairs (excitons) are found to be the elementary quasiparticle excitation of the theory.

Above discussed counterintutive features of our theory can be understood by noting that 
the Lagrange multiplier field introduced to implement current constraint (equation (\ref{z2})), minimally couples to Dirac fermions 
and hence manifests like an effective magnetic field. However, the sign of this magnetic field is opposite for both the valley fermions,
which now move in mutually opposite cyclotron orbits in accordance to effective magnetic field felt by them.  
As shown in Fig. \ref{fig}, the current constraint is obeyed at any point in bulk, as the contribution to local current coming 
from cyclotron loops from both valley fermions would be equal. At Graphene boundary, the contribution from cyclotron loops present 
in vicinity adds up, giving rise to clockwise and counter clockwise circulating edge currents of equal magnitude. 
Classically(at tree level), these two currents add up to zero and hence no edge current exist. However, in one dimensional theories like Graphene boundary, quantum corrections generate terms at one loop level, which violate conservation of chiral currents. As a result of this,
the two opposite current flows no longer cancel each other and a net nonzero edge contribution survives. Fermions at the two valleys 
living in bulk feel exactly opposite effective magnetic field,  so presence of an external magnetic field creates an asymmetry 
between the two valley fermions, which now feel different magnetic fields. It is this difference of magnetic fields that results in
difference between cyclotron orbits amongst two valleys, and leads to non zero circulating current in the bulk. This ultimately gives rise
to net diamagnetic response of the system. A neutral composite particle like exciton would not couple to any of these fields, 
and hence would be a long lived quasiparticle excitation in absence of Dirac fermion, which does not show up in the Hilbert space of asymptotic fields.

Above theory is reminiscent of a model of superconductivity, proposed by Laughlin \cite{laughlin}, which goes by the name of 
anyon superconductivity \cite{hosotani,fetter,chen,panigrahi} (for a lucid introduction and broad review see \cite{wilczekbook}). It was shown in this context that, fermions interacting with an emergent statistical gauge field can become anyons, and show superconductivity, albeit without any pairing. A crucial feature in which our theory differs from anyon superconductivity is that, in the latter case time reversal symmetry is explicitly broken due to presence of Chern-Simons term in action, which is responsible for fractional (anyonic) spin and statistics of quasiparticles. On the otherhand, in the present proposal, time reversal symmetry is preserved and quasiparticles do not have anomalous spin/statistics. Although above theory has strong resemblance with the one proposed in Ref. \cite{shreecharan},  it needs to be pointed out, that these two theories are fundamentally different, in the sense that in the latter case one does have Dirac fermion as a propagating mode, whereas in present case one does not.

\noindent \textit{Acknowledgements-} The authors acknowledge useful discussions with Prof. Jainendra Jain and Dr. S. Lal regarding this work.



%% References


\bibliographystyle{model1a-num-names}
\bibliography{constraint.bib}






% \end{thebibliography}

\end{document}